\documentclass{article}
\usepackage{spconf,amsmath,graphicx, siunitx, graphicx,subcaption,multicol, tabularx, multirow, array}
\usepackage{adjustbox}

\title{The secret source : Incorporating source features to improve Acoustic-to-articulatory speech inversion}
\name{Yashish M. Siriwardena and Carol Espy-Wilson \thanks{© 2022 IEEE. Personal use of this material is permitted. Permission from IEEE must be obtained for all other uses, in any current or future media, including reprinting/republishing this material for advertising or promotional purposes, creating new collective works, for resale or redistribution to servers or lists, or reuse of any copyrighted component of this work in other works.}
\thanks{This work was supported by the National Science Foundation grant IIS1764010.}}

\address{Institute for Systems Research, University of Maryland College Park, USA \\}

\begin{document}
\ninept
\vspace{-5pt}
\maketitle
\vspace{-5pt}
\begin{abstract}

In this work, we incorporated acoustically derived source features, aperiodicity, periodicity and pitch as additional targets to an acoustic-to-articulatory speech inversion (SI) system. We also propose a Temporal Convolution based SI system, which uses auditory spectrograms as the input speech representation, to learn long-range dependencies and complex interactions between the source and vocal tract, to improve the SI task. The experiments are conducted with both the Wisconsin X-ray microbeam (XRMB) and Haskins Production Rate Comparison (HPRC) datasets, with comparisons done with respect to three baseline SI model architectures. The proposed SI system with the HPRC dataset gains an improvement of close to 28\% when the source features are used as additional targets. The same SI system outperforms the current best performing SI models by around 9\% on the XRMB dataset. 

\end{abstract}
%
\vspace*{-8pt}
\section{Introduction}
\label{sec:intro}
\vspace{-5pt}

Speech production is a complex process involving coordinated movements of the articulators: lips, jaw, tongue, teeth, glottis and soft palate. The vibration of vocal folds at the glottis or the lack of it determines the periodicity of the generated sound. Apart from the vocal fold vibration, aspiration, frication and transients are identified as other sound sources \cite{stevens2000acoustic}. The vocal tract which consists of the velum, tongue, lips and teeth, acts like an acoustic tube which modulates the source waveform. The acoustics of all vowel productions and most of the consonants have been described by a linear source-filter theory \cite{stevens2000acoustic}. This theory is based on the assumption that the source of speech production is independent of the vocal tract filter. However, the actual process of speech production is nonlinear since the aerodynamics inside the glottis and vocal tract is governed by non-linear equations, and most importantly it has been shown that there exists a mutual interaction between the source and filter in certain cases of speech production \cite{Titze2008-ac}. In this work, we take into account this source-filter interaction to improve the acoustic-to-articulatory speech inversion task.

The inverse problem of determining the trajectories of the movement of speech articulators from the speech signal is referred to as acoustic-to-articulatory speech inversion \cite{Sivaraman_ASA, SI_Trans_ppr}. Learning this mapping from acoustics to articulation is an ill-posed problem which is known to be highly non-linear and non-unique \cite{Qin2007}. Speaker variability in speech production makes it even harder or if not impossible to develop speaker-independent SI systems. However, accurate estimation of articulatory trajectories or vocal tract variables can benefit speech applications like Automatic Speech Recognition (ASR) \cite{Mitra2010}, speech synthesis \cite{speech_synthesis_1}, speech therapy \cite{Fagel2008A3V} and mental health assessments \cite{espywilson19_interspeech, Siriwardena_SZ}. Ground-truth articulatory data are collected by techniques like X-ray microbeam \cite{Westbury1994a}, Electromagnetic Articulometry (EMA) \cite{Tiede2017} and real-time Magnetic Resonance Imaging (rt-MRI) \cite{Narayanan2004}. All these methods are expensive, time consuming and need specialized equipment for observing articulatory movements directly \cite{Sivaraman_ASA}. For these reasons, developing a speaker-independent SI system that can accurately estimate articulatory features for any unseen speaker can potentially transform how speech research is conducted. 

Recent advancements in deep neural networks (DNNs) and learning algorithms have significantly contributed to improving SI systems over the last few years. Bidirectional LSTMs (BiLSTMS) \cite{illa18_interspeech}, BiGRNNs \cite{yashish_bigrnn}, CNN-BiLSTMs \cite{Shahrebabaki2020}, Temporal Convolutional Networks (TCN) \cite{shahrebabaki21_interspeech} and transformer models \cite{udupa21_interspeech} have gained state-of-the-art results with multiple articulatory datasets. Most of these SI systems use either extracted acoustic features like Mel Frequency Cepstral Coefficients (MFCCs), Mel-spectrograms or the waveform itself as the input speech representation, and learns a mapping to the ground-truth articulatory variables. To further improve the SI task, phoneme features have been used as inputs along with the acoustic features in \cite{Shahrebabaki2020}. One limitation of these models is that you need phonetic transcriptions of the speech utterances at the time of inference. To address this issue while also leveraging additional information that phonetic transcriptions offer, multi task learning frameworks were proposed in \cite{siriwardena22_interspeech} where phoneme labels are jointly predicted with TVs as targets. But, to the best of our knowledge, source level features have not been explored either as inputs or as targets to improve the SI task. One main reason for that can be the lack of ground-truth source data (like Electroglottography (EGG) recorded synchronously with EMA or XRMB recordings) that can be directly used to capture the glottal activity in the existing articulatory datasets.  Hence, the current state of the art SI systems trained with these existing articulatory datasets are usually not able to estimate any of the glottal (or velar) activity in speech. 

In this work, we investigated the idea of estimating source level features along with the TVs as targets to leverage any source-filter interactions to improve the overall SI task. Here we used source features from the Aperiodicity, Periodicity and Pitch (APP) detector \cite{APPdetector} as proxies for the source activity. We also experimented with multiple input representations of speech with different DNN model architectures and used two publicly available articulatory datasets to show the significance of learning source level information in improving the acoustic-to-articulatory speech inversion task.





\vspace*{-8pt}
\section{Articulatory Datasets}
\vspace*{-2pt}
\label{sec:dataset}

\subsection{X-Ray Microbeam (XRMB) dataset}
\vspace*{-2pt}
\label{ssec:xrmb_dataset}

The original University of Wisconsin XRMB database \cite{Westbury1994b} comprises of naturally spoken isolated sentences and short read paragraphs collected from 32 male and 25 female subjects. These speech utterances were recorded along with trajectory data captured by X-ray microbeam cinematography of the midsagittal plane of the vocal tract using pellets placed on several articulators: upper (UL) and lower (LL) lip, tongue tip (T1), tongue blade (T2), tongue dorsum (T3), tongue root (T4), mandible incisor (MANi), and (parasagittally placed) mandible molar (MANm). However, some of the articulatory recordings were marked as mistracked in the database and eliminating these samples left us with 46 speakers (21 males and 25 females) with a total of around 4 hours of speech data.

For each of the above mentioned articulators, the X-Y positions of the pallet movement is recorded. Since the X-Y positions of the pellets strongly depend on the anatomy of the speakers and variability of pellet placements, the measurements can vary significantly across speakers. Hence, to better represent vocal tract shape, relative measures were used to calculate the Tract Variables (TVs) from the X-Y positions of the pellets. TVs lead to a relatively speaker independent representation of speech articulation and characterize salient features of the vocal tract area function \cite{McGowan1994}. The TVs are based on articulatory phonology, a theoretical framework for speech production \cite{Browman1992}. Using geometric transformations, the XRMB trajectories were converted to TV trajectories as outlined in \cite{Mitra2012}. The transformed XRMB database comprises of six TV trajectories: Lip Aperture (LA), Lip Protrusion (LP), Tongue Body Constriction Location (TBCL), Tongue Body Constriction Degree (TBCD), Tongue Tip Constriction Location (TTCL) and, Tongue Tip Constriction Degree (TTCD).  

\vspace*{-8pt}
\subsection{Haskins Production Rate Comparison (HPRC) dataset}
\vspace*{-2pt}
\label{ssec:xrmb_dataset}

We also used the HPRC database which contains recordings from 4 female and 4 male subjects reciting 720 phonetically balanced IEEE sentences \cite{IEEE_sentences} at normal and fast production rates \cite{Tiede2017}. The recordings were done using a 5-D electromagnetic articulometry (EMA) system (WAVE; Northern Digital). First, every sentence was produced at the speaker’s preferred ‘normal’ speaking rate and then a ‘fast’ repetition of the same, without making errors. Sensors were placed on the tongue (tip (TT), body (TB), root (TR)), lips (upper (UL) and lower (LL)) and mandible, together with reference sensors on the left and right mastoids, and upper and lower incisors (UI, LI). These EMA trajectories were obtained at 100 Hz and the synchronized audio were recorded at 44.1 KHz. Geometric transformations as defined in \cite{Sivaraman_ASA} were used to obtain 9 TVs (namely  Lip Aperture (LA), Lip Protrusion (LP), Tongue Body Constriction Location (TBCL), Tongue Body Constriction Degree (TBCD), Tongue Tip Constriction Location (TTCL), Tongue Tip Constriction Degree (TTCD), Jaw Angle (JA), Tongue Middle Constriction Location (TMCL) and Tongue Middle Constriction Degree (TMCD)).

\begin{figure}[th]
    \centering
    \includegraphics[width=\linewidth, height=45mm]{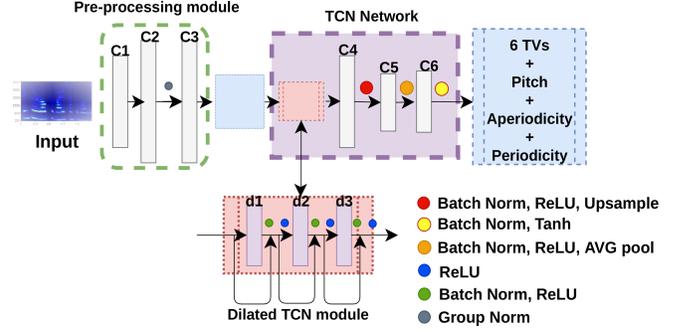}
    \caption{Model architecture of the SI system. Here C1-C6 represent 1D-CNN layers where as d1-d3 represent 1D dilated CNN layers}
    \label{fig:model_archi}
\end{figure}

\begin{table*}[t]
    \centering
    \footnotesize
    \caption{PPMC scores for articulatory variable prediction on the XRMB dataset. Model names with `SF' uses source features as additional targets. The AVG. TVs column for those models also show the percentage increase in TV prediction with respect to the same model which does not use source features}
    \vspace{-8pt}
    \label{tab:ppmc_scores_xrmb}
    \begin{tabular}{|l|l|l|l|l|l|l|l|l|l|l|l|}
    \hline
    \textbf{Model}     & \textbf{LA} &\textbf{LP} &\textbf{TBCL} & \textbf{TBCD} &\textbf{TTCL} &\textbf{TTCD} & \textbf{Ap.} &\textbf{Per.} &\textbf{Pitch} &\textbf{AVG. TVs} &\textbf{AVG. all}\\ \hline
    TCN-Audspec  &0.7977 &0.7942 &0.7883 &0.7836 &0.7743  &0.7684 &-  &- &- &0.7844 &-\\ \hline
    TCN-SF-Audspec &\textbf{0.8448}  &\textbf{0.8640} &\textbf{0.8604} &\textbf{0.8818} &\textbf{0.9029} &\textbf{0.9005} &\textbf{0.9082} &\textbf{0.8860} &\textbf{0.9021} &\textbf{0.8770} (9.3\%) &\textbf{0.8834} \\\hline
    TCN-Mspec  &0.7432 &0.7427 &0.7366 &0.7244 &0.7179 &0.6993 &- &- &- &0.7273 &- \\ \hline
    TCN-SF-Mspec  &0.8364 &0.8639 &0.8727 &0.8607 &0.8807 &0.8917 &0.8732 &0.9005 &0.8638 &0.8677 (14\%) &0.8715 \\ \hline
    BiGRNN-MFCC  &0.8801 &0.6200 &0.8580 &0.7382 &0.6922 &0.9206 &- &- &- &0.7848 &- \\ \hline
    BiGRNN-SF-MFCC  &0.8810 &0.6211 &0.8628 &0.7365 &0.7019 &0.9191 &0.8693 &0.9163 &0.7209 &0.7871 (0.2\%) &0.8032 \\ \hline
    CNN-BiGRNN-Mspec  &0.8801 &0.6165 &0.8505 &0.7355 &0.7146 &0.9171 &- &- &- &0.7858 &- \\ \hline
    CNN-BiGRNN-SF-Mspec  &0.8799 &0.6246 &0.8566 &0.7302 &0.7065 &0.9175 &0.8794 &0.9296 &0.7441 &0.7859 (0.01\%) &0.8076 \\ \hline
    CNN-BLSTM-Mspec  &0.8770 &0.6184 &0.8463 &0.7200 &0.6915 &0.9197 &- &- &- &0.7788 &- \\ \hline
    CNN-BLSTM-SF-Mspec  &0.8774 &0.6202 &0.8525 &0.7172 &0.6941 &0.9180 &0.8734 &0.9263 &0.7442 &0.7799 (0.1\%) &0.8026 \\ \hline
    \end{tabular}
\end{table*}

\section{Proposed speech inversion system}
\vspace*{-2pt}
\label{sec:dataset}

\vspace*{-5pt}
\subsection{Input representations}
\label{ssec:model_impl}
\vspace*{-2pt}

All the audio files are first segmented into 2second long segments and the shorter audios are zero padded at the end. However for the HPRC dataset, the audio files are first down-sampled to 16 KHz before segmentation and padding. The following features are then extracted from the segmented audio utterances.

\vspace*{-8pt}
\subsubsection{Auditory Spectrograms}
\label{ssec:model_impl}
\vspace*{-2pt}

We converted the one-dimensional pressure time waveform into a two-dimensional pattern of neural activity distributed along the tonotopic axis (roughly a logarithmic frequency). This two-dimensional representation, which is defined as an `auditory spectrogram' (Audspec) \cite{Wang_auditory_spec} is used in our proposed SI system as the input speech representation. It has been shown that this Audspec is an enhanced and a noise-robust estimate of the Fourier-based spectrogram \cite{Wang_auditory_spec}. 

\vspace*{-8pt}
\subsubsection{MFCCs and Melspectrorgams}
\label{ssec:mfccs_melspcs}
\vspace*{-2pt}

Mel-Frequency Cepstral Coefficients (MFCCs) and Melspectrograms (MSPECs) are extracted as the acoustic input features for baseline SI systems. Both MFCCs and MSPECs were extracted using a 20ms Hamming analysis window with a 10ms frame shift. For MFCCs, 13 cepstral coefficients were extracted for each frame while 40 Mel frequencies were used for both MFCCs and MSPECs. Both MFCCs and MSPECs are utterance wise normalized (z-normalized) prior to model training. 


\vspace*{-5pt}
\subsection{Model Architecture and training}
\label{ssec:model_impl}
\vspace*{-2pt}

We propose a Temporal Convolution Network (TCN) based acoustic-to-articulatory SI system which takes in the Audspecs as input. The proposed system, unlike conventional SI systems, estimates both TVs and source level features (aperiodicity, periodicity and pitch) as the output. The model is optimized using the Mean Squared Error (MSE) loss computed between the predicted articulatory variables and the ground truth (TVs from articulatory datasets and source features from APP detector \cite{APPdetector}). 

The SI system is implemented in PyTorch with 1-D convolutional (CNN) layers. The complete network is inspired by the multilayered Temporal Convolution Network in \cite{Lea_2017_Temporal_CNN}. Figure \ref{fig:model_archi} shows the proposed model architecture with its sub-modules used for pre processing and dilated TCN. The Pre-processing module contains three 1-D CNN layers with 1$\times$1 kernels (C1, C2 and C3), which have 128, 256 and 256 filters, respectively. The d1, d2 and d3 dilated CNN layers have a kernel size of 3 with 1,4 and 16 dilation rates respectively. Upsampling (window size 4) is done after C4 layer and average pooling (window size 5) is done after C5 layer along with BatchNorm layers after every CNN layer in the TCN network. The upsampling and average pooling operations take care of matching the time dimension of the input spectrograms to the target time dimension of TVs.

To train the SI system, learning rates were determined based on a grid search by testing all combinations from [1e-2, 1e-3, 1e-4, 3e-4] that resulted in 1e-3 as the best pick. A similar grid search was done to choose the batch size from [16, 32, 64, 128] and 64 gave the best validation MSE. The objective function was optimized using the ADAM optimizer with an `ExponentialLR' learning rate scheduler and a decay of 0.5. All models were trained by monitoring the validation loss.

To train the models with the XRMB dataset, the dataset was divided into training, development, and testing splits, so that the training set has utterances from 36 speakers and the development and testing sets have 5 speakers each (3 males,2 females). To train the models with the HPRC dataset, similar to the XRMB dataset, the dataset was divided into training, development, and testing sets, so that the training set has utterances from 6 speakers (3 Males, 3 Females) and the development and testing sets have utterances of 2 speakers (1 male,1 female) equally split between them. We used audio samples with both the normal and fast production rates in the HPRC dataset. To create the validation and test sets, the audio samples from the two speakers were randomly split, so that both the splits have samples corresponding to normal and fast rates.

For both the datasets, none of the training splits have overlapping speakers with the development and testing sets and hence all the models are trained in a `speaker-independent' fashion. The splits also ensured that around 80\% of the total number of utterances were present in training, and the development and testing sets have a nearly equal number of utterances. This allocation was done in a completely random manner.

\vspace*{-10pt}
\subsection{Performance metric}
\label{ssec:model_impl}
\vspace*{-2pt}

All the models are evaluated with the Pearson Product Moment Correlation (PPMC) scores computed between the estimated articulatory variables and the corresponding ground-truth. Equation \ref{eq1} is used to computed the PPMC score, where $X$ represents the estimated articulatory variable, $\overline{X}$ the mean of the estimated, $Y$ the ground-truth variable, $\overline{Y}$ the mean of the ground-truth variable and $N$ the number of articulatory variables.

\vspace*{-5pt}
\begin{equation}
  PPMC = \frac{\sum_i^N{(X[i] - \overline{X})(Y[i] - \overline{Y})}}{\sqrt{\sum_i^N{(X[i] - \overline{X})^2(Y[i] - \overline{Y})^2}}}
  \label{eq1}
\end{equation}
\vspace*{-5pt}

\begin{table}[t]
    \centering
    \normalsize
    \caption{PPMC scores for HPRC dataset.}
    \vspace{-8pt}
    \label{tab:ppmc_scores_hprc}
    \begin{tabular}{|l|l|l|}
    \hline
    \textbf{Model} &\textbf{AVG. 9 TVs} &\textbf{AVG. all}\\ \hline
    TCN-Audspec &0.4805 &-\\ \hline
    TCN-SF-Audspec &\textbf{0.7573} (27.7\%) &\textbf{0.7636} \\\hline
    TCN-Mspec &0.4763 &- \\ \hline
    TCN-SF-Mspec &0.6503 (17.4\%) &0.6621 \\ \hline
    BiGRNN-MFCC &0.7118 &- \\ \hline
    BiGRNN-SF-MFCC &0.7153 (0.3\%) &0.7263 \\ \hline
    CNN-BiGRNN-Mspec &0.7277 &- \\ \hline
    CNN-BiGRNN-SF-Mspec &0.7290 (0.1\%) &0.7461 \\ \hline
    CNN-BLSTM-Mspec &0.7245 &- \\ \hline
    CNN-BLSTM-SF-Mspec &0.7259 (0.1\%) &0.7428 \\ \hline
    \end{tabular}
\end{table}

\vspace*{-5pt}
\section{Baseline Speech Inversion Systems}
\vspace*{-5pt}
\label{sec:baseline_models}

This section discusses about the baseline SI systems implemented for comparison. Detailed information on the model architectures and implementation can be found in a GitHub repository \footnote[1]{https://github.com/Yashish92/Speech-Inversion-TCN}

\vspace*{-10pt}
\subsection{BiGRNN model}
\label{ssec:model_bigrnn}
\vspace*{-2pt}

We used the BiGRNN model architecture implemented in \cite{yashish_bigrnn, siriwardena22_interspeech} as one of our baseline models for comparison. The model has 2 bidirectional layers of Gated Recurrent Units (GRUs) followed by two time distributed fully connected layers. Dropout layers are also used after every layer to minimize the issue of over-fitting. 13 MFCCs (extracted as in section \ref{ssec:mfccs_melspcs}) are used as input to this SI system.

\vspace*{-10pt}
\subsection{CNN-BiLSTM and CNN-BiGRNN model}
\label{ssec:model_cnn-bilstm}
\vspace*{-2pt}

A CNN-BiLSTM model inspired by the work in \cite{Shahrebabaki2020} was implemented as another baseline SI system. The model consists of 5 CNN (1D CNNs) layers, whose outputs are then concatenated together and fed to 2 BiLSTM layers. The output from the last BiLSTM layer is then passed through two time distributed fully connected layers, where the final fully connected layer serves as the output layer. 

A similar architecture was used to implement a CNN-BiGRNN model where the only difference is that the 2 BiLSTM layers in the CNN-BiLSTM model are now replaced with bidirectional layers of Gated Recurrent Units (GRUs). CNN-BiGRNN model is comparatively light weight due to the GRUs used in the model instead of LSTM units. A BiGRNN based SI system has also shown to outperform a conventional BiLSTM based SI system in \cite{yashish_bigrnn} which motivated this new CNN-BiGRNN model as a baseline for comparison.  

Similar to the BiGRNN model in section \ref{ssec:model_bigrnn}, dropout layers are used after every layer to minimize possible over-fitting in both the CNN-BiLSTM and CNN-BiGRNN models. Both models used MSPECs as the input speech representation.

\vspace*{-5pt}
\section{Results}
\vspace*{-2pt}
\label{sec:dataset}  

\subsection{Comparison with baseline SI systems}
\label{ssec:model_impl}
\vspace*{-2pt}

The baseline models discussed in section \ref{sec:baseline_models} were trained and evaluated with the same train-dev-test splits for comparison. For every model architecture, two versions of the model were implemented with one only predicting the TVs as targets (6 TVs for XRMB dataset, 9 TVs for HPRC dataset) and the other predicting both TVs and source features. 

A baseline TCN based SI system (TCN-Audspec) was trained with Audspecs as input and only TVs as targets for both XRMB and HPRC datasets. A similar TCN architecture was implemented to use MSPECs as inputs and, two versions of this model (TCN-Mspec and TCN-SF-Mspec) were trained similar to the other baseline models. Table \ref{tab:ppmc_scores_xrmb} shows the PPMC scores for TV estimation on the XRMB dataset. The PPMC scores for individual TVs and source features are listed here along with average scores across TVs and all the predicted articulatory variables (TVs + source features). Similarly, Table \ref{tab:ppmc_scores_hprc} lists the average PPMC scores across the 9 TVs and all the articulatory variables for the HPRC dataset. 


\vspace*{-9pt}
\subsection{Estimated TVs and source features}
\label{ssec:model_impl}
\vspace*{-2pt}

\begin{figure}[th]
    \centering
    \includegraphics[width=\linewidth,  height=75mm]{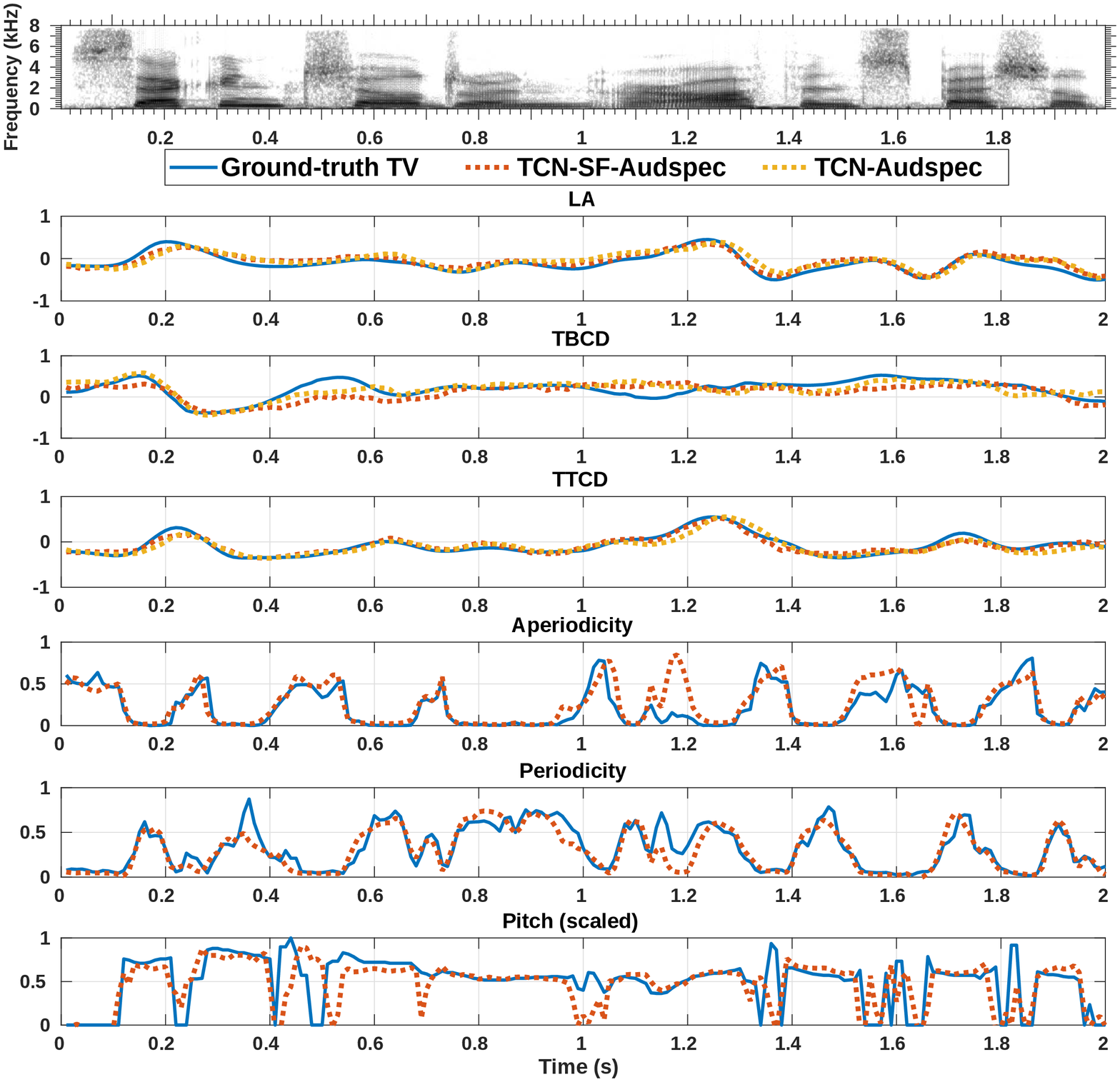}
    \caption{LA and constriction degree TVs + source features for the utterance ‘second children are often special’ estimated by the proposed TCN-SF-Audspec model compared to the TCN-Audspec. Solid blue Line - ground truth, red dotted line - predictions by the TCN-SF-Audspec, yellow dotted Line - predictions by TCN-Audspec.}
    \label{fig:tv_tracks}
\end{figure}

Figure \ref{fig:tv_tracks} shows the estimated constriction degree TVs (LA, TBCD, TTCD) and source features from the proposed TCN-SF-Audspec and the TCN-Audspec models. As can be observed in the plots, the source features are predicted with a considerably better accuracy, which is an added advantage of the proposed SI system. This also gives an almost complete articulatory representation of speech (only missing velar activity) that can be useful in various speech applications (e.g articulatory speech synthesis).

\vspace*{-12pt}
\section{Discussion}
\label{sec:discus}
\vspace{-2pt} 

Deep Neural Networks have been effective in learning complex non-linear relationships between input speech representations and articulatory movements that has lead to the recent success in acoustic-to-articulatory SI task. However, there is still a lot of room for improvement, especially in developing speaker-independent and more generalizable SI systems that can be effectively utilized in speech applications. To this end, we explored the idea of incorporating source characteristics to exploit any source-filter interactions, and thereby mutually learn, and improve, the acoustic-to-articluatory SI task. 

We used aperiodicity, periodicity and pitch as source features, which are used as additional targets to train the SI systems. For the proposed TCN based SI system and for every baseline model, two versions of the model were trained with the goal of investigating the real effect of incorporating source features. As shown in Tables \ref{tab:ppmc_scores_xrmb} and \ref{tab:ppmc_scores_hprc}, the results are consistent across both articulatory datasets, and support the fact that incorporating source features into the mix of TVs is definitely helping the estimation of articulatory variables. This observation is quite evident with the proposed TCN models which use Audspecs or MSPECs as inputs. For example, with the XRMB dataset, the TCN model which uses MSEPCs gain an absolute improvement of around 14\% with respect to the same model which does not use source features as targets. Similarly with the HPRC dataset, the TCN model which uses Audspecs gain an absolute improvement of close to 28\% when the source features are used as additional targets. Most importantly, when the best PPMC scores for average TV estimation with the proposed TCN model is considered, it is around a 9\% improvement over the current best performing SI systems in \cite{Sivaraman_ASA, yashish_bigrnn}, trained and evaluated on the same splits of the XRMB dataset. 

However, a key observation here is that both the input speech representation and the DNN model architecture play a significant role in learning these complex relationship between the source features and TVs. For example, with the 13 MFCCs, which is the most commonly used speech representation in SI systems, adding source features as targets does not significantly improve the PPMC scores. This is consistent with both the XRMB and HPRC datasets with the BiGRNN model which uses 13 MFCCs as input. This can be mainly due to the fact that the 13 MFCCs do not contain important source information and is usually limited to capturing the filter characteristics (vocal tract) in speech production. Moreover, having richer speech representations that contain source information does not necessarily mean it will improve on the SI task. A fine observation to support that is the CNN-BiGRNN and CNN-BiLSTM models that use MSPECs as inputs, which necessarily contain valuable source information unlike the 13 MFCCs. This elucidate the fact that the DNN model architecture too plays a critical role in learning these complex dependencies between the source and articulatory targets.

TCN based models have shown to be extremely effective in speech applications \cite{TCN_enhancement} with learning long-range temporal (and contextual) dependencies. These models have shown to outperform typical RNN and CNN based models, especially in applications where subtle and complex contextual information needs to be extracted from input representations \cite{Lea_2017_Temporal_CNN}. This was one of the key motivations for the proposed SI system, which ultimately outperformed all the other baseline systems on both XRMB and HPRC datasets. However, as mentioned earlier, the input representation used to train these models also play a role which can be clearly observed with results in Tables \ref{tab:ppmc_scores_xrmb} and \ref{tab:ppmc_scores_hprc}. The TCN model trained with Audspecs as input is outperforming the same TCN based model architecture trained with MSPECs which suggests that the Audspecs might be capturing important spectral and temporal information that is helpful in mutually learning both the source features and TVs.  

Conventional SI systems usually predict constriction degree related TVs significantly better (eg., LA, TTCD) with respect to the constriction location related TVs (eg., LP, TTCL). The same can be more or less observed (with the exception of TBCL and TBCD) for individual TV predictions in Table \ref{tab:ppmc_scores_xrmb} for all the `baseline models'. Surprisingly, this is not as evident with the TCN based models which tend to predict both the location and degree TVs with close to similar accuracies. Moreover, further analysis needs to be done to investigate the ways and instances by which the source features are actually interacting with the TVs, and also to understand what the TCN models are actually capturing as source-filter interactions that is ultimately helping the overall SI task.








\bibliographystyle{IEEEbib}
\bibliography{mybib}

\begin{thebibliography}{10}

\bibitem{stevens2000acoustic}
Kenneth~N Stevens,
\newblock {\em Acoustic phonetics}, vol.~30,
\newblock MIT press, 2000.

\bibitem{Titze2008-ac}
Ingo Titze, Tobias Riede, and Peter Popolo,
\newblock ``Nonlinear source-filter coupling in phonation: vocal exercises,''
\newblock {\em J Acoust Soc Am}, vol. 123, no. 4, pp. 1902--1915, Apr. 2008.

\bibitem{Sivaraman_ASA}
Ganesh Sivaraman, Vikramjit Mitra, Hosung Nam, Mark Tiede, and Carol
  Espy-Wilson,
\newblock ``Unsupervised speaker adaptation for speaker independent acoustic to
  articulatory speech inversion,''
\newblock {\em The Journal of the Acoustical Society of America}, vol. 146, no.
  1, pp. 316--329, 2019.

\bibitem{SI_Trans_ppr}
Abdolreza~Sabzi Shahrebabaki, Giampiero Salvi, Torbjørn Svendsen, and
  Sabato~Marco Siniscalchi,
\newblock ``Acoustic-to-articulatory mapping with joint optimization of deep
  speech enhancement and articulatory inversion models,''
\newblock {\em IEEE/ACM Transactions on Audio, Speech, and Language
  Processing}, vol. 30, pp. 135--147, 2022.

\bibitem{Qin2007}
Chao Qin and Miguel~{\'{A}} Carreira-Perpi{\~{n}}{\'{a}}n,
\newblock ``{An empirical investigation of the nonuniqueness in the
  acoustic-to-articulatory mapping.},''
\newblock {\em Interspeech}, pp. 74--77, 2007.

\bibitem{Mitra2010}
Vikramjit Mitra, Hosung Nam, Carol~Y. Espy-Wilson, Elliot Saltzman, and Louis
  Goldstein,
\newblock ``{Retrieving tract variables from acoustics: A comparison of
  different machine learning strategies},''
\newblock {\em IEEE Journal on Selected Topics in Signal Processing}, vol. 4,
  no. 6, pp. 1027--1045, sep 2010.

\bibitem{speech_synthesis_1}
Zhen-Hua Ling, Korin Richmond, and Junichi Yamagishi,
\newblock ``Articulatory control of hmm-based parametric speech synthesis using
  feature-space-switched multiple regression,''
\newblock {\em IEEE Transactions on Audio, Speech, and Language Processing},
  vol. 21, no. 1, pp. 207--219, 2013.

\bibitem{Fagel2008A3V}
Sascha Fagel and Katja Madany,
\newblock ``A 3-d virtual head as a tool for speech therapy for children,''
\newblock in {\em INTERSPEECH}, 2008.

\bibitem{espywilson19_interspeech}
Carol Espy-Wilson, Adam~C. Lammert, Nadee Seneviratne, and Thomas~F. Quatieri,
\newblock ``{Assessing Neuromotor Coordination in Depression Using Inverted
  Vocal Tract Variables},''
\newblock in {\em Proc. Interspeech 2019}, 2019, pp. 1448--1452.

\bibitem{Siriwardena_SZ}
Yashish~M. Siriwardena, Carol Espy-Wilson, Chris Kitchen, and Deanna~L. Kelly,
\newblock {\em Multimodal Approach for Assessing Neuromotor Coordination in
  Schizophrenia Using Convolutional Neural Networks}, p. 768–772,
\newblock Association for Computing Machinery, New York, NY, USA, 2021.

\bibitem{Westbury1994a}
John~R Westbury,
\newblock ``{Speech Production Database User ' S Handbook},''
\newblock {\em IEEE Personal Communications - IEEE Pers. Commun.}, vol. 0, no.
  June, 1994.

\bibitem{Tiede2017}
Mark Tiede, Carol~Y. Espy-Wilson, Dolly Goldenberg, Vikramjit Mitra, Hosung
  Nam, and Ganesh Sivaraman,
\newblock ``Quantifying kinematic aspects of reduction in a contrasting rate
  production task,''
\newblock {\em The Journal of the Acoustical Society of America}, vol. 141, no.
  5, pp. 3580--3580, 2017.

\bibitem{Narayanan2004}
Shrikanth Narayanan, Krishna Nayak, Sungbok Lee, Abhinav Sethy, and Dani Byrd,
\newblock ``{An approach to real-time magnetic resonance imaging for speech
  production},''
\newblock {\em The Journal of the Acoustical Society of America}, vol. 115, no.
  4, pp. 1771--1776, mar 2004.

\bibitem{illa18_interspeech}
Aravind Illa and Prasanta~Kumar Ghosh,
\newblock ``{Low Resource Acoustic-to-articulatory Inversion Using
  Bi-directional Long Short Term Memory},''
\newblock in {\em Proc. Interspeech 2018}, 2018, pp. 3122--3126.

\bibitem{yashish_bigrnn}
Yashish~M. Siriwardena, Ahmed~Adel Attia, Ganesh Sivaraman, and Carol
  Espy-Wilson,
\newblock ``Audio data augmentation for acoustic-to-articulatory speech
  inversion using bidirectional gated rnns,''
\newblock arXiv 2022.

\bibitem{Shahrebabaki2020}
Abdolreza~Sabzi Shahrebabaki, Sabato~Marco Siniscalchi, Giampiero Salvi, and
  Torbjørn Svendsen,
\newblock ``{Sequence-to-Sequence Articulatory Inversion Through Time
  Convolution of Sub-Band Frequency Signals},''
\newblock in {\em Proc. Interspeech 2020}, 2020, pp. 2882--2886.

\bibitem{shahrebabaki21_interspeech}
Abdolreza~Sabzi Shahrebabaki, Sabato~Marco Siniscalchi, and Torbjørn Svendsen,
\newblock ``{Raw Speech-to-Articulatory Inversion by Temporal Filtering and
  Decimation},''
\newblock in {\em Proc. Interspeech 2021}, 2021, pp. 1184--1188.

\bibitem{udupa21_interspeech}
Sathvik Udupa, Anwesha Roy, Abhayjeet Singh, Aravind Illa, and Prasanta~Kumar
  Ghosh,
\newblock ``{Estimating Articulatory Movements in Speech Production with
  Transformer Networks},''
\newblock in {\em Proc. Interspeech 2021}, 2021, pp. 1154--1158.

\bibitem{siriwardena22_interspeech}
Yashish~M. Siriwardena, Ganesh Sivaraman, and Carol Espy-Wilson,
\newblock ``{Acoustic-to-articulatory Speech Inversion with Multi-task
  Learning},''
\newblock in {\em Proc. Interspeech 2022}, 2022, pp. 5020--5024.

\bibitem{APPdetector}
O.~{Deshmukh}, C.~Y. {Espy-Wilson}, A.~{Salomon}, and J.~{Singh},
\newblock ``Use of temporal information: detection of periodicity,
  aperiodicity, and pitch in speech,''
\newblock {\em IEEE Transactions on Speech and Audio Processing}, vol. 13, no.
  5, pp. 776--786, 2005.

\bibitem{Westbury1994b}
John~R Westbury,
\newblock ``{Speech Production Database User's Handbook},''
\newblock {\em IEEE Personal Communications}, vol. 0, no. June, 1994.

\bibitem{McGowan1994}
Richard~S. McGowan,
\newblock ``{Recovering articulatory movement from formant frequency
  trajectories using task dynamics and a genetic algorithm: Preliminary model
  tests},''
\newblock {\em Speech Communication}, vol. 14, no. 1, pp. 19--48, 1994.

\bibitem{Browman1992}
Catherine~P Browman and Louis Goldstein,
\newblock ``{Articulatory Phonology : An Overview *},''
\newblock {\em Phonetica}, vol. 49, pp. 155--180, 1992.

\bibitem{Mitra2012}
Vikramjit Mitra, Hosung Nam, Carol Espy-Wilson, Elliot Saltzman, and Louis
  Goldstein,
\newblock ``{Recognizing articulatory gestures from speech for robust speech
  recognition},''
\newblock {\em The Journal of the Acoustical Society of America}, vol. 131, no.
  3, pp. 2270--2287, 2012.

\bibitem{IEEE_sentences}
``Ieee recommended practice for speech quality measurements,''
\newblock {\em IEEE Transactions on Audio and Electroacoustics}, vol. 17, no.
  3, pp. 225--246, 1969.

\bibitem{Wang_auditory_spec}
Kuansan Wang and S.~Shamma,
\newblock ``Self-normalization and noise-robustness in early auditory
  representations,''
\newblock {\em IEEE Transactions on Speech and Audio Processing}, vol. 2, no.
  3, pp. 421--435, 1994.

\bibitem{Lea_2017_Temporal_CNN}
Colin Lea, Michael~D. Flynn, Rene Vidal, Austin Reiter, and Gregory~D. Hager,
\newblock ``Temporal convolutional networks for action segmentation and
  detection,''
\newblock in {\em Proceedings of the IEEE Conference on Computer Vision and
  Pattern Recognition (CVPR)}, July 2017.

\bibitem{TCN_enhancement}
Ashutosh Pandey and DeLiang Wang,
\newblock ``Tcnn: Temporal convolutional neural network for real-time speech
  enhancement in the time domain,''
\newblock in {\em ICASSP 2019 - 2019 IEEE International Conference on
  Acoustics, Speech and Signal Processing (ICASSP)}, 2019, pp. 6875--6879.

\end{thebibliography}

\end{document}